\documentclass[%
aps,
prl,%
amsmath,amssymb,showpacs,
reprint,%
floatfix,
twocolumn,
nofootinbib,
longbibliography,
superscriptaddress
]{revtex4-2}
\pdfoutput=0

\usepackage[]{todonotes}
\usepackage{amsmath,amsfonts,amssymb}

\usepackage{hyperref}

\def\doubleunderline#1{\underline{\underline{#1}}} 

\newcommand{\levshift}{{\delta}}
\newcommand{\omegazero}{\omega_\mathrm{a}^{(0)}}

\allowdisplaybreaks[3]

\makeatletter
\newcommand{\raisemath}[1]{\mathpalette{\raisem@th{#1}}}
\newcommand{\raisem@th}[3]{\raisebox{#1}{$#2#3$}}
\makeatother

\usepackage{makecell}
\usepackage{pbox}

\usepackage{float}

\usepackage{xr}
\usepackage[markup=red,deletedmarkup=red]{changes}
\definechangesauthor[name=Jerg, color=red]{JE}
\definechangesauthor[name=Dominik, color=magenta]{DL}

\newcommand{\dl}[1]{{\color{black}#1}}

\begin{document}
	
	\title{Certifying multi-mode light-matter interaction in lossy resonators}
	
	\author{Dominik \surname{Lentrodt}}
	\email[]{dominik.lentrodt@mpi-hd.mpg.de}
	
	\affiliation{Max-Planck-Institut f\"ur Kernphysik, Saupfercheckweg 1, 69117 Heidelberg, Germany}
	\affiliation{Physikalisches Institut, Albert-Ludwigs-Universit\"at Freiburg, Hermann-Herder-Stra{\ss}e 3, D-79104 Freiburg, Germany}
	\affiliation{EUCOR Centre for Quantum Science and Quantum Computing, Albert-Ludwigs-Universit\"at Freiburg, Hermann-Herder-Stra{\ss}e 3, D-79104 Freiburg, Germany}

	\author{Oliver~\surname{Diekmann}}
	
	\affiliation{Max-Planck-Institut f\"ur Kernphysik, Saupfercheckweg 1, 69117 Heidelberg, Germany}
	\affiliation{Institute for Theoretical Physics, Vienna University of Technology (TU Wien), 1040 Vienna, Austria}
	
	\author{Christoph H.~\surname{Keitel}}
	
	\affiliation{Max-Planck-Institut f\"ur Kernphysik, Saupfercheckweg 1, 69117 Heidelberg, Germany}
	
	\author{Stefan~\surname{Rotter}}
	
	\affiliation{Institute for Theoretical Physics, Vienna University of Technology (TU Wien), 1040 Vienna, Austria}
	
	\author{J\"org \surname{Evers}}
	\email[]{joerg.evers@mpi-hd.mpg.de}
	
    \affiliation{Max-Planck-Institut f\"ur Kernphysik, Saupfercheckweg 1, 69117 Heidelberg, Germany}
	
	\begin{abstract}
		Quantum models based on few-mode Master equations have been a central tool in the study of resonator quantum electrodynamics, extending the seminal single-mode Jaynes-Cummings model to include loss and multiple modes. Despite their broad application range, previous approaches within this framework have either relied on a Markov approximation or a fitting procedure. By combining ideas from pseudomode and quasinormal mode theory, \dl{we develop a certification criterion for multi-mode effects in lossy resonators. It is based on a witness observable, and neither requires a fitting procedure nor a Markov approximation.} Using the resulting criterion, we demonstrate that such multi-mode effects are important for understanding previous experiments in X-ray cavity QED with M\"ossbauer nuclei and that they allow one to tune the nuclear ensemble properties.
	\end{abstract}
	
	\maketitle
	
	In the study of quantum matter interacting with light fields, an important theoretical \cite{Jaynes1963a,Kirton2018} and experimental paradigm \cite{Haroche2006,Kavokin2017_BOOK} is the coupling to a single cavity mode~\cite{Krimer2014a}.
	More recently, regimes beyond single-mode light-matter interaction have moved into the focus of attention. In particular, multi-mode \dl{resonator quantum electrodynamics} has opened new possibilities for many-body quantum systems \cite{Ritsch2013}, for example by enabling the tuning of effective interactions \cite{Vaidya2018}\dl{, and for quantum information processing \cite{Blais2021,Chakram2021}.}
	Also for single atoms, qualitatively new dynamics can be realized when multiple modes participate, such as in the multi-mode strong coupling regime \cite{Krimer2014}, where the extreme light-matter coupling bridges the free spectral range between the modes~\cite{Johnson2019,Blaha2022}.
	
	For many platforms, the realization of extreme coupling strengths or the required mode control remain technically challenging. However, multi-mode features may also appear in the opposite regime of open and absorptive resonators, e.g., if the mode widths exceed their frequency spacing~\cite{Hackenbroich2002,Stockmann2002,Lalanne2018}. Interestingly, strong losses do not necessarily represent an obstacle, but may instead cause fascinating phenomena, such as for lasers \cite{Hodaei2014,Peng2014a,Miao2016} and in the context of non-Hermitian physics \cite{Dittes2000,El-Ganainy2018,Miri2019,Feng2017,Longhi2017,Ashida2020_review}. Within quantum optics and particularly nanophotonics, where complex mode structures can be engineered \cite{Peng2017} and losses are often sizable \cite{Liu2017}, non-Hermitian effects have attracted considerable interest, with potential applications ranging from quantum plasmonics \cite{Tame2013} to cavity-controlled materials \cite{GarciaVidal2021,Schlawin2022} and chemistry \cite{Ribeiro2018,Feist2018,Thomas2019,Ulusoy2020a,Schafer2022,Li2022}.
		
	Another example is X-ray cavity QED (cQED) with ensembles of M\"ossbauer  nuclei~\cite{Rohlsberger2021}. It uniquely combines the possibility of high-precision X-ray spectroscopy using the extreme quality factors of the involved nuclear resonances with cavities which are restricted to the lossy regime due to the low refractive index contrast at hard X-ray energies. As a result, single-mode models fail to quantitatively describe the nuclear spectra \cite{Rohlsberger2010,Heeg2015a} and standard fitting procedures for the multi-mode case require heuristic extensions~\cite{Heeg2015c}. This regime is of practical importance, since such cavity-nuclei systems facilitate the engineering of artificial few-level quantum systems, which are otherwise inaccessible at hard X-ray energies~\cite{Rohlsberger2010,Rohlsberger2012,Heeg2013a,Heeg2015b,Haber2016a,Haber2017,Heeg2013b,Longo2016,Lentrodt2020a,Diekmann2022,Diekmann2022b}.
	
	\dl{From a theory perspective, these regimes of cQED are particularly challenging, due to an interplay of strong light-matter coupling and large cavity losses.} The latter implies a breakdown of standard Jaynes-Cummings (JC) models \cite{Barnett1988a,Hackenbroich2002,Fregoni2022}, which feature a few discrete cavity modes coupled to an external bath \cite{Gardiner1985,Gardiner2004}. \dl{Similar issues with such models are encountered at ultra-strong coupling \cite{Beaudoin2011,Bamba2014,SanchezMunoz2018,FriskKockum2019}.} \dl{While many alternative approaches to describe non-Markovian open quantum systems exist \cite{Breuer2002_BOOK,DeVega2017}, including various types of Master equations only involving the matter degrees of freedom \cite{Nakajima1958,Zwanzig1960,Redfield1965,Shibata1977}, \dl{quantum stochastic approaches \cite{Carmichael1999,Gardiner2004,Tanimura2020}}, and direct treatments of the continuum \cite{Krimer2014,Roy2017,Malekakhlagh2017,Strathearn2018,SanchezBarquilla2020}, such methods remain challenging particularly for complex matter systems.} As a result, much effort has been invested in finding suitable generalizations of and a rigorous theoretical basis for few-mode models.
		
	 \dl{Current approaches to this problem fall into three distinct categories.} On the one hand, the recent quantization of quasinormal modes \cite{Franke2019} has promoted techniques from semi-classical and perturbative light-matter interaction theory \cite{Lalanne2018,Kristensen2020} to the quantum level. While this approach allows for insights into absorptive resonator physics \cite{Franke2020a,Franke2020b,Franke2021,Franke2022,Ren2021,Ren2022}, its applicability is limited by the use of Markov approximations~\cite{Franke2019_supplement}, which break down \dl{for} high loss \dl{regimes} \cite{Gardiner1985,Hackenbroich2002}. Similarly, \textit{ab initio} system-bath theory \cite{Gardiner1985,Viviescas2003} allows one to derive exact few-mode Hamiltonians \cite{Lentrodt2020}, but generally requires a Markov approximation when transferring to a Master equation. On the other hand, the concept of pseudomodes \cite{Garraway1997a,Dalton2001,Tamascelli2018,Trivedi2021}, \dl{its non-Hermitian variants \cite{Lamprecht1999,Lamprecht2002,Pleasance2020,Pleasance2021_arxiv}}, and the related chain mappings \cite{Denning2019,Trivedi2021,SanchezBarquilla2021,Nuesseler2022} from the theory of open quantum systems are designed to \dl{reduce the continuum bath to discrete modes with} minimal or no Markov approximation at all. \dl{This idea can be seen as a special discrete mode basis \cite{Tamascelli2018,Medina2021}} where the \dl{system-bath coupling \cite{Gardiner1985,Dalton2001,Hackenbroich2002,Lentrodt2020} is frequency-independent}. However, current methods to construct such exactly Markovian descriptions rely on \dl{fitting procedures to match the bath properties \cite{Mascherpa2020}, such as the spectral density in the case of cQED~\cite{Medina2021}.}
	
	In this Letter, we combine these previous approaches to develop a certification criterion for multi-mode quantum effects that neither relies on a Markov approximation nor a fitting procedure.

	Our method employs pseudomodes-based few-mode (pFM) models, which do not require a Markov approximation, but in general involve a priori unknown parameters. \dl{Therefore, we construct a witness observable} to certify the necessity of multiple modes and their interactions in the pFM model, and to classify different cases of multi-mode effects. \dl{By establishing a connection to classical quasinormal mode (cQNM) expansions of the witness observable}, which is related to the Green's function \cite{Lalanne2018,Kristensen2014} characterizing the electromagnetic environment of a quantum emitter, the need for fitting model parameters is eliminated.
	Within X-ray cQED, we apply our certification criterion to unambiguously identify the previously measured collective nuclear Lamb shift \cite{Rohlsberger2010} as a multi-mode effect, resolving a puzzle raised by earlier models \cite{Heeg2013b,Heeg2015c}. We further show how the different types of multi-mode effects allow one to invert the sign of this shift by engineering suitable cavity environments.
	
\begin{figure}[t]
		\includegraphics[width=1.0\columnwidth,height=6cm]{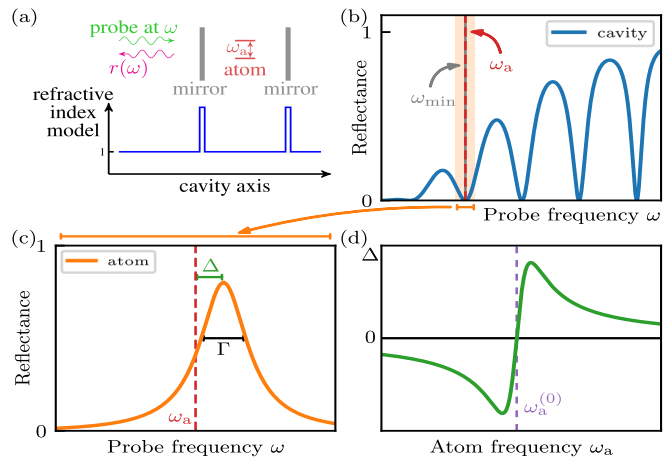}
		\caption{Model setup and the spectroscopic quantities considered in the analysis. (a) Model cavity coupled to a two-level atom at its center. (b) Reflectance for the case of spectrally broad cavity modes and a narrow atomic resonance. (c) shows a magnification of (b) around the atomic resonance.
		(d) $\Delta$ as a function of the atomic transition frequency. \label{fig::illu_new1}}
	\end{figure}

	\textit{Revisiting the single-mode case.---}We start by discussing features of a standard single-mode pFM model of a generic cavity, see Fig.~\ref{fig::illu_new1}(a). Without atom, the frequency-dependent reflection coefficient is~\cite{supplement_mmcls2021}
	\begin{align}\label{eq::emptyCavitySpectrum}
	r_\mathrm{cav}(\omega) = 1 - 2\pi i \frac{|\kappa_R|^2}{\omega - \omega_1 + i\frac{\kappa}{2}} \,,
	\end{align}
	where $\omega$ is the probing light frequency, $\kappa_R$ the coupling rate in and out of the cavity mode of frequency $\omega_1$, and $\kappa$ the total cavity loss rate.  We denote the relevant minimum of this empty-cavity reflectance ($|r_\mathrm{cav}|^2$) as $\omega_\mathrm{min}$, which in the single-mode case coincides with $\omega_1$. An example reflectance (featuring multiple modes) is shown in Fig.~\ref{fig::illu_new1}(b).
	
	Coupling a single two-level atom with rate $g$ to the cavity mode leads to a lossy JC model~\cite{Jaynes1963a,Haroche2006}. \dl{The} linear reflection spectrum is a Lorentzian on the cavity background~\cite{Nunnenkamp2011,Lentrodt2020,supplement_mmcls2021}
	\begin{align}\label{eq::singleMode_refSpec}
		r(\omega) = r_\mathrm{cav}(\omega) - 2\pi i  \frac{(\kappa^{(\mathrm{int})}_R\dl{(\omega)})^2}{\omega - \omega_\textrm{a} - \Delta\dl{(\omega)} + i\Gamma\dl{(\omega)}/2} \,,
	\end{align}
	\dl{where the Lorentzian parameters are frequency dependent to ensure applicability at strong coupling.} \dl{Specifically, $(\kappa^{(\mathrm{int})}_R\dl{(\omega)})^2 = |\kappa_Rg|^2/(\omega - \omega_1 + i\kappa/2)^2$ and the real quantities $\Delta(\omega)$ and $\Gamma(\omega)$ together define the function
	\begin{align}\label{eq::single_mode_parameters}
		\levshift(\omega) :=\Delta(\omega) - i \frac{\Gamma(\omega)}{2} = \frac{gg^*}{\omega - \omega_1 + i\frac{\kappa}{2}} \,,
	\end{align}
	which can be extracted from the reflection spectrum [Fig.~\ref{fig::illu_new1}(c)]. At weak coupling, $\Delta=\Delta(\omega_\mathrm{a})$, $\Gamma=\Gamma(\omega_\mathrm{a})$ correspond to the Lamb shift and Purcell enhanced line width of the atom, respectively. Beyond weak coupling, these quantities are closely related to bath correlation functions, e.g.~$\Gamma(\omega)$ is proportional to the spectral density \cite{Medina2021}.}
	As a function of the atomic transition frequency $\omega_\textrm{a}$, the single-mode case results in a dispersion-like shape for $\Delta$ [Fig.~\ref{fig::illu_new1}(d)].  We denote the atomic resonance frequency at which $\Delta=0$ as $\omegazero$.
	
	By inspection of Eqs.~(\ref{eq::emptyCavitySpectrum})-(\ref{eq::single_mode_parameters}), we find that the minimum  of the empty cavity reflectance $\omega_\mathrm{min}$, the zero of the frequency shift $\Delta(\omegazero)=0$, the maximum of the line width broadening $\Gamma$, and the real part $\omega_1$ of the pole location of $\Delta(\omega)$ in the complex frequency plane all coincide at the cavity mode frequency [$\omega_\mathrm{min} = \omegazero = \omega_1$]. As one consequence, single-mode cavity models do not allow for non-zero shifts $\Delta$ at the minimum $\omega_\mathrm{min}$ of the cavity background $r_\mathrm{cav}$, which is at odds with the experimental observation of such a non-zero shift reported in Fig.~3 of \cite{Rohlsberger2010}. \dl{Motivated by this, we focus the following discussion on the Lamb shift $\Delta$ as a \textit{witness observable}, which we demonstrate to provide diagnostic information about multi-mode effects.}
	\begin{figure}[t]
		\includegraphics[width=1.0\columnwidth]{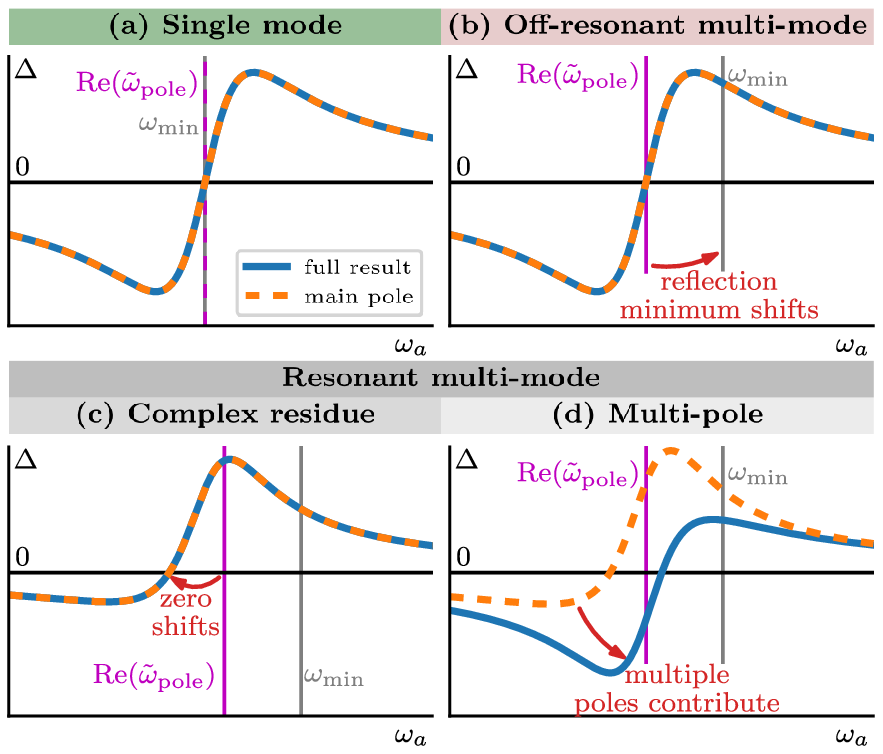}
		\includegraphics[width=1.0\columnwidth]{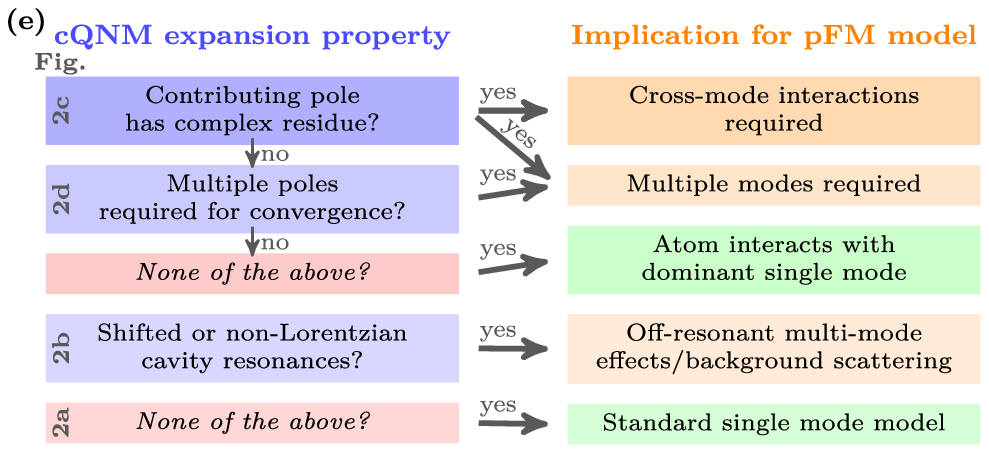}
		\caption{Certification and classification of multi-mode effects, illustrated via the Lamb shift $\Delta$ as in Fig.~\ref{fig::illu_new1}(d). (a) Single-mode case with its feature coincidence (see text). (b) Off-resonant multi-mode effects can appear already in the empty-cavity reflectance, and lift the degeneracy of $\mathrm{Re}(\tilde{\omega}_\mathrm{pole})$ and $\omega_\mathrm{min}$. (c, d) Resonant multi-mode effects affecting the light-matter interaction. (c) In the complex-residue case, the main pole residue has an imaginary contribution, unlike in the single-mode case. (d) In the multi-pole case, more than one pole is required for convergence. \dl{(e) Decision tree for certifying multi-mode effects using the cQNM expansion of the witness observable.} \label{fig::illu_new2}}
	\end{figure}

	\textit{Multi-mode case.}---Next, we generalize to the multi-mode case. We employ a pFM model of a recently developed form \cite{Medina2021,supplement_mmcls2021}, which applies to complex and absorptive resonators \dl{described by macroscopic QED} \cite{Scheel2008}. It is given by the Master equation~\cite{Medina2021}
	\begin{align}\label{eq::mm_master}
		\dot{\rho} = -i[H; \rho] + \sum_i \frac{\kappa_i}{2} \left(2 \hat{a}^{\mathstrut}_i \rho \hat{a}_i^\dagger - \hat{a}_i^\dagger \hat{a}^{\mathstrut}_i \rho - \rho \hat{a}_i^\dagger\hat{a}^{\mathstrut}_i\right) \,,
	\end{align}
	where $H=H_\mathrm{cav} + H_\mathrm{atom}$, with an interacting cavity mode Hamiltonian 
	$H_\mathrm{cav} = \sum_{ij} \omega_{ij} \hat{a}_i^\dagger \hat{a}^{\mathstrut}_j$, and the JC interaction $H_\mathrm{atom} = (\omega_\mathrm{a}/2)\hat{\sigma}^z + \sum_i \left[g^*_i  \hat{a}^{\mathstrut}_i \hat{\sigma}^+ + h.c.\right]$.
	Eq.~\eqref{eq::single_mode_parameters} then generalizes to \cite{Lentrodt2020,supplement_mmcls2021}
	\begin{align}
		\levshift &= \underline{g}^\dagger\, \left[\omega_\mathrm{a}-\doubleunderline{\omega}_{\raisemath{2pt}{\mathrm{cav}}} + \frac{i}{2}\doubleunderline{\kappa}\,\right]^{-1} \,\underline{g}\,, \label{eq:levshift-mode}
	\end{align}
	where $(\underline{g})_i=g_i$, $(\doubleunderline{\omega}_{\raisemath{2pt}{\mathrm{cav}}})_{ij} = \omega_{ij}$ and $(\doubleunderline{\kappa})_{ij} = \kappa_i\delta_{ij}$ are the multi-mode couplings, cavity mode interactions and decay terms, respectively, in matrix-vector notation.
	
	Importantly, despite the simple form of Eq.~(\ref{eq::mm_master}), this model does not employ a Markov approximation \cite{Medina2021} due to its use of the pseudomodes concept \cite{Garraway1997b,Tamascelli2018,Pleasance2020}. \dl{In contrast, the quantized QNM approach in \cite{Franke2019} requires an explicit Markov and related approximations \cite{Franke2019_supplement}, such that the resulting Master equations are not necessarily the same as pFM models}. \dl{However, obtaining the parameters of the latter currently requires fitting procedures~\cite{Mascherpa2020,Medina2021}, even in the ideal case when the cavity geometry is known without experimental uncertainties.}
	
	\textit{From few-mode models to quasinormal mode expansions.}---
	In order to certify and categorize multi-mode effects in the pFM approach, we diagonalize Eq.~\eqref{eq:levshift-mode} as
	\begin{align}\label{eq::ML_poleExpansion_final_mainText}
		\levshift = \sum_i \frac{\bar{g}_i^*\tilde{g}_i}{\omega_\mathrm{a} - \tilde{\Omega}_i + i \frac{\tilde{\kappa}_i}{2}} \,,
	\end{align}
	where the diagonal basis parameters relate to the bare mode parameters in Eq.~\eqref{eq:levshift-mode} by an invertible transformation matrix \cite{supplement_mmcls2021,Stockmann2002}. Note that here we assumed that the diagonalization exists and comprises only simple poles. A more general treatment is required, e.g., in the presence of exceptional points~\cite{Miri2019,El-Ganainy2018}.

	In order not to rely on a fit, we use an alternative expression from cQNM theory for the level shift in terms of the cavity's classical Green's function $\mathbf{G}$, given by
	 $\levshift = -\mu_0 \omega_\mathrm{a}^2 \mathbf{d}^* \cdot \mathbf{G}(\mathbf{r}_\mathrm{a}, \mathbf{r}_\mathrm{a}, \omega_\mathrm{a}) \cdot \mathbf{d}$~\cite{Dung2000,Scheel2008,Buhmann2008,Asenjo-Garcia2017a} where $\mathbf{d}$ is the transition's dipole moment vector. $\mathbf{G}$ can then be expanded using a pole expansion~\cite{Lalanne2018,Defrance2020,Wu2021a,supplement_mmcls2021}, such that
	 \begin{align}\label{eq::ML_poleExpansion_final_physical}
		\levshift  = \sum_i \frac{r_i}{\omega_\mathrm{a} - \tilde{\omega}_{\mathrm{pole},i}} \,.
	\end{align}
	\dl{Such expansions are known as quasinormal mode expansions in the electromagnetism literature, which have been studied extensively \cite{Lalanne2018,Kristensen2020} and are of the same form as the diagonalized pFM expansion Eq.~(\ref{eq::ML_poleExpansion_final_mainText}). We can therefore use these cQNM expansions of the witness observable as an intermediate step to draw conclusions about the pFM model.}
	
	\begin{figure}[t]
		\includegraphics[width=1.0\columnwidth]{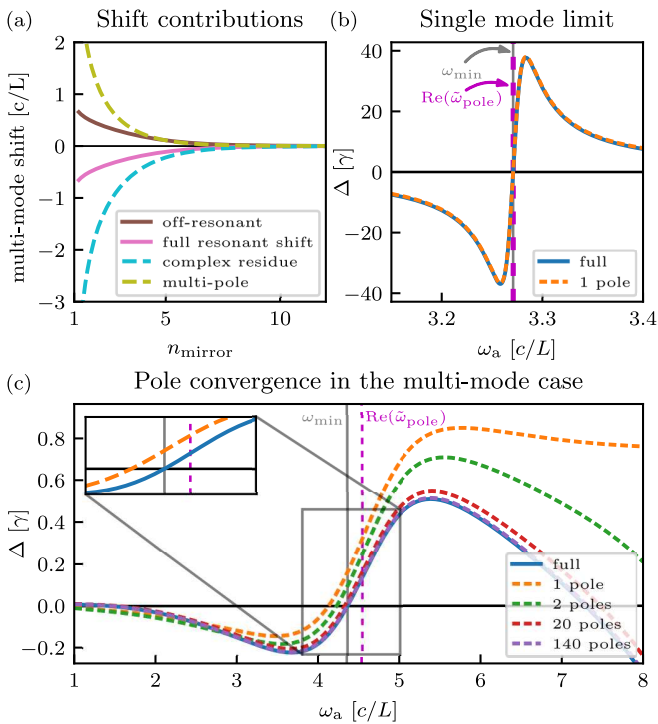}
		\caption{Transition from the single-mode to the multi-mode case in Fabry-P\'erot-like cavities. Panel (a) quantifies the shift contribution of each multi-mode effect to the zero position of $\Delta(\omega_\mathrm{a})$ (see legend and text).
		(b) Cavity-induced energy shift $\Delta$ in the single-mode limit ($n_\mathrm{mirror} = 20$). (c) Corresponding results in a multi-mode case ($n_\mathrm{mirror} = 4$), where multiple poles are required to achieve convergence, and even the single-pole line shape differs from the single-mode case due to complex residue effects. The inset shows the full and single-pole results around the pole location to make their shifts visible. $\gamma$ is the free space radiative line width. \label{fig::cls_mm_shift_new}}
	\end{figure}
	
	\textit{Certification of multi-mode effects.}---
	This comparison leads to three distinct cases of multi-mode effects, which deviate in different ways from the single-mode formula Eq.~\eqref{eq::single_mode_parameters}. We illustrate each case by its action on the witness observable $\Delta(\omega_\textrm{a})$ (see Fig.~\ref{fig::illu_new2}) \dl{and implications for the resulting pFM Master equation.}

    In the fully single-mode case [Fig.~\ref{fig::illu_new2}(a)], already the main pole provides a good quantitative description, with the aforementioned frequency coincidence $\omegazero=\mathrm{Re}(\tilde{\omega}_\mathrm{pole})=\omega_\mathrm{min}$.
	
	The first condition for the single-mode case is thus that a single main pole has to be sufficient to achieve convergence of the pole expansion in a relevant frequency range. The opposite case, where multiple poles are required for good convergence, we will refer to as \textit{multi-pole effects} [Fig.~\ref{fig::illu_new2}(d)]. Multiple poles in the cQNM expansion thus also imply that multiple modes are needed in the pFM model.
	
	Second, we also require real-valued residues, since the residue $|g|^2$ in the single-mode case Eq.~(\ref{eq::single_mode_parameters}) is real. \textit{Complex-residue effects} appear if a pole has a non-negligible imaginary part of its residue \cite{Denning2019,Cernotik2019,Pleasance2021_arxiv}. Then, already the first pole contribution to the shape of $\Delta(\omega_\textrm{a})$ may be modified and its zero $\omegazero$ moves away from the pole location $\mathrm{Re}(\tilde{\omega}_\mathrm{pole})$ [Fig.~\ref{fig::illu_new2}(c)]. Since an isolated pFM mode always features a Lorentzian with a real residue, the presence of complex residues is a direct indicator of \textit{cross-mode interactions} \cite{Hackenbroich2002,Viviescas2003,Medina2021}\dl{, unless non-Hermitian interaction Hamiltonians are employed \cite{Lamprecht1999,Lamprecht2002,Pleasance2021_arxiv}. Even the case of single QNM physics may therefore require multiple modes in the pFM theory, implying that the two are distinct concepts.}
	
	Both complex-residue and multi-pole effects are \textit{resonant effects}, since they directly affect the cavity-modified properties $\Delta$ and $\Gamma$ of the atom, and \textit{multi-mode effects}, since any Markovian single-mode model fails to capture them. In addition, single-mode models may fail even in the absence of the atom, if the empty-cavity reflectance deviates from its single-mode form Eq.~\eqref{eq::emptyCavitySpectrum} due to spectral overlap between cavity modes. These \textit{off-resonant multi-mode effects} typically shift the reflectance minimum $\omega_\mathrm{min}$ away from the pole and the zero of $\Delta$ \cite{Heeg2015c} [Fig.~\ref{fig::illu_new2}(b)]. \dl{On the level of the pFM model, these off-resonant effects can also be understood as a manifestation of background scattering contributions in input-output theory~\cite{Lentrodt2020}. Similar issues are encountered in cQNM theory, where various approaches to expand the scattering matrix have been investigated recently \cite{Alpeggiani2017,Ren2020,Binkowski2020,Benzaouia2021}.}
	
	Analyzing the properties of the cQNM expansion therefore allows to certify and distinguish multi-mode effects in pFM models [see Fig.~\ref{fig::illu_new2}(e)]. We note that while our criterion focuses on certifying multi-mode effects resulting from large losses and overlapping modes, the pFM Master equation fully applies at strong coupling~\cite{Lentrodt2020,Medina2021,supplement_mmcls2021}. \dl{\dl{The witness observable $\Delta$ can be interpreted as the line shift at weak coupling, but also has physical meaning beyond this regime \cite{Krimer2014}.}}
	
	\textit{Multi-mode effects in Fabry-P\'erot-like cavities.}---
	Next, we explore the relevance of the different multi-mode effects by analyzing a simple model reminiscent of archetype Fabry-P\'erot cavities [Fig.~\ref{fig::illu_new1}(a)]. The cavity of length $L$ is modeled via a piece-wise constant one-dimensional refractive index comprising vacuum with two mirror layers of refractive index $n_\mathrm{mirror}$ and thickness $t_\mathrm{mirror}=L/100$. This model allows us to study the continuous transition between the single and multi-mode pFM case by inducing spectral mode overlap via high leakage through the mirrors.
	
	Results are shown in Fig.~\ref{fig::cls_mm_shift_new}.  Panel (a) quantifies resonant multi-mode effects via the displacement of the zero $\omega_\textrm{a}^{(0)}$ of $\Delta(\omega_\textrm{a})$ from the main cQNM pole position $\mathrm{Re}(\tilde{\omega}_\mathrm{pole})$ as a function of $n_\mathrm{mirror}$, and off-resonant multi-mode effects via the shift of  $\mathrm{Re}(\tilde{\omega}_\mathrm{pole})$ from $\omega_\mathrm{min}$. For the single-mode case in high-quality resonators with well-isolated resonances ($n_\mathrm{mirror}\gg 1$), all shifts vanish. As expected, towards lower-quality resonators, the multi-mode effects become increasingly relevant. Interestingly, the large negative complex-residue shift is partially canceled by the multi-pole and off-resonant shifts in this cavity. Panels (b,c) show $\Delta$ as a function of the atomic transition frequency $\omega_\textrm{a}$ for a single-mode (b, $n_\mathrm{mirror} = 20$) and a multi-mode (c, $n_\mathrm{mirror} = 4$) scenario. In the pFM single-mode case, one cQNM pole is sufficient to model the system, and no zero-shift of $\Delta$ appears. In contrast, the multi-mode case in (c) requires many poles for convergence. The single-pole approximation further has a significantly complex residue. These observations are incompatible with single-mode models according to Eqs.~(\ref{eq::emptyCavitySpectrum})-(\ref{eq::single_mode_parameters}).s

	\begin{figure}[t]
		\includegraphics[width=\columnwidth]{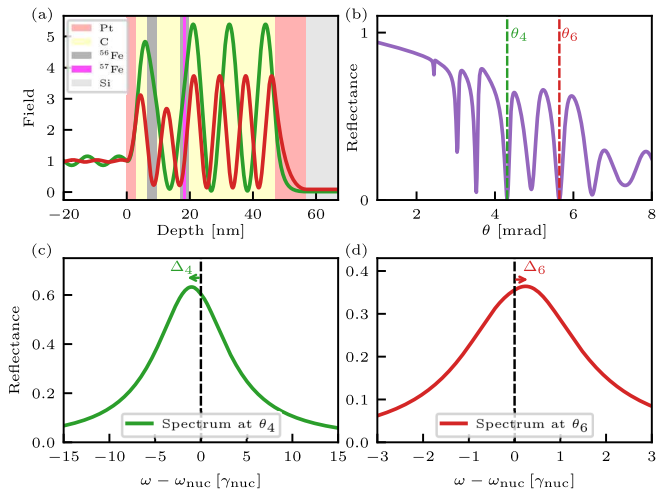}
		\caption{Design of multi-mode level shifts. 
		(a) Cavity geometry and off-resonant field distributions in the fourth ($\theta_4$, green) and sixth ($\theta_6$, red) cavity mode. The cavity structure is Pt (3~nm)/C (3.5~nm)/$^{56}$Fe (3~nm)/C (7.5~nm)/$^{56}$Fe (1~nm)/$^{57}$Fe (1~nm)/$^{56}$Fe (1~nm)/C (27~nm)/Pt (10~nm)/Si.
		(b) Off-resonant cavity reflectance as function of incidence angle showing the overlapping mode structure. (c,d) Nuclear spectra at $\theta_4, \theta_6$, demonstrating the reversed energy shift in the two critically coupled mode minima. \label{fig::cls_inversion}}
	\end{figure}
	
	\textit{Multi-mode level shift design.}---Finally, we explore the multi-mode effects outlined above in the concrete platform of thin-film X-ray cQED with M\"ossbauer nuclei, which has recently attracted considerable theoretical \cite{Heeg2013b,Heeg2015c,Heeg2016arxiv,Kong2016,Lentrodt2020a,Kong2020} and experimental \cite{Rohlsberger2010,Rohlsberger2012,Heeg2013a,Heeg2015a,Heeg2015b,Haber2016a,Haber2017} attention. Similar setups with electronic processes are starting to be explored \cite{Haber2019,Gu2021,Vassholz2021}. In these cavities, the atom is replaced by a thin layer containing an ensemble of M\"ossbauer nuclei. 
	Because of the low refractive-index contrast, the cavities are probed at grazing incidence and generally feature large losses and overlapping modes \cite{Heeg2015c}. In the experimentally relevant low-excitation regime, the cavity-nuclei system can be approximated as a few-level quantum system \cite{Heeg2013b,Lentrodt2020a,Andrejic2021,supplement_mmcls2021}, comprising the collective ground state and excitonic single-excitation states coherently spread over the nuclei. The latter couple via transverse cavity modes at a given parallel wave vector~\cite{Lentrodt2020a}.
	Such cavities thus form a promising platform to implement quantum optical schemes at hard X-ray energies, which are otherwise unavailable due to the lack of suitable X-ray driving fields. For this purpose, the possibility to control the quantum optical parameters is key to implement more advanced quantum optical level schemes~\cite{Longo2016,Rohlsberger2021}. From an experimental point of view, operation at a critically coupled minimum of the empty-cavity reflectance $\omega_\mathrm{min}$ is highly desirable, in order to suppress the spectrally broad electronic scattering background. In this configuration, however, the single-mode case severely impedes the parameter control due to the strict coincidence of spectral features discussed earlier.
	
	The calculations in Fig.~\ref{fig::cls_inversion} show that the above-developed multi-mode effects provide a physical mechanism to overcome this challenge. The cavity structure shown in panel (a) features overlapping cavity resonances visible in the off-resonant cavity reflectance shown in (b). Panels (c,d) show the resulting spectra of the nuclear resonances for driving at the two modes $\theta_4$ and $\theta_6$, respectively, corresponding to criticially coupled minima. The cavity structure is designed such that the multi-mode effects lead to opposite signs of the respective collective nuclear Lamb shifts $\Delta$. This sign flip demonstrates that tuning the mode environment through the cavity geometry allows one to deliberately alter the effective level scheme of the nuclear ensemble in a qualitative way, as it is required, e.g., to satisfy delicate detuning conditions in more complex multi-level systems \cite{Fleischhauer2005,Rohlsberger2012,Heeg2015c}.
	A closer inspection~\cite{supplement_mmcls2021} shows that the fourth mode in this cavity mainly features off-resonant multi-mode effects, while the sixth mode is influenced by resonant multi-mode effects. An analogous analysis \cite{supplement_mmcls2021} reveals that already the first experimental observation of the collective Lamb shift in a similar cavity \cite{Rohlsberger2010} relies on an off-resonant multi-mode effect, illustrating the importance for the interpretation of spectroscopic observables, which has been a puzzle within previous models \cite{Heeg2013b,Heeg2015c}. We note that while the resonance shifts are smaller than the superradiant line width, they are measurable \cite{Rohlsberger2010,Heeg2015a} and understanding whether they arise due to cavity or material effects is important for the interpretation of experimental spectra, and for the design of more advanced quantum optical level schemes using tailored cavity reservoirs~\cite{Diekmann2022,Diekmann2022b}\dl{, where generalized witness observables may be constructed}. More pronounced signatures of such multi-mode effects appear in multi-layer systems \cite{Rohlsberger2012}, where the spectral interference strongly depends on multi-mode contributions~\cite{Heeg2015c}.
	
	\begin{acknowledgments}
		The authors would like to thank L.~Bocklage, M.~W.~Haverkort and K.~P.~Heeg for valuable discussions. DL gratefully acknowledges the Georg
		H. Endress Foundation for financial support and support by the DFG funded Research Training Group ``Dynamics of Controlled Atomic and Molecular Systems'' (RTG 2717). OD gratefully acknowledges financial support by the Cusanuswerk, the Studienstiftung des Deutschen Volkes and the Austrian Science Fund (FWF) Grant No.~P32300. The code used in this paper can be accessed at~\cite{Lentrodt2022_Zenodo1} and an open-source \textsc{python} library for parts of these computations has been made available at~\cite{Lentrodt2022_pyrot}.
	\end{acknowledgments}
	
	\bibliographystyle{myprsty}
	\bibliography{library}

\end{document}


\title{Certifying multi-mode light-matter interaction in lossy resonators\\Supplemental Material}

	\author{Dominik \surname{Lentrodt}}
	\email[]{dominik.lentrodt@mpi-hd.mpg.de}
	
	\affiliation{Max-Planck-Institut f\"ur Kernphysik, Saupfercheckweg 1, 69117 Heidelberg, Germany}
	\affiliation{Physikalisches Institut, Albert-Ludwigs-Universit\"at Freiburg, Hermann-Herder-Stra{\ss}e 3, D-79104 Freiburg, Germany}
	\affiliation{EUCOR Centre for Quantum Science and Quantum Computing, Albert-Ludwigs-Universit\"at Freiburg, Hermann-Herder-Stra{\ss}e 3, D-79104 Freiburg, Germany}

	\author{Oliver~\surname{Diekmann}}
	
	\affiliation{Max-Planck-Institut f\"ur Kernphysik, Saupfercheckweg 1, 69117 Heidelberg, Germany}
	\affiliation{Institute for Theoretical Physics, Vienna University of Technology (TU Wien), 1040 Vienna, Austria}
	
	\author{Christoph H.~\surname{Keitel}}
	
	\affiliation{Max-Planck-Institut f\"ur Kernphysik, Saupfercheckweg 1, 69117 Heidelberg, Germany}
	
	\author{Stefan~\surname{Rotter}}
	
	\affiliation{Institute for Theoretical Physics, Vienna University of Technology (TU Wien), 1040 Vienna, Austria}
	
	\author{J\"org \surname{Evers}}
	\email[]{joerg.evers@mpi-hd.mpg.de}
	
    \affiliation{Max-Planck-Institut f\"ur Kernphysik, Saupfercheckweg 1, 69117 Heidelberg, Germany}

	\maketitle

	This supplemental material provides details on the open single-mode Jaynes-Cummings model, the construction of pseudomodes-based few-mode (pFM) models and resulting expansion of the complex level shift, its relation to the quasinormal mode expansion, and further analyzes the multi-mode effects in the x-ray cavity QED example in the main text.

	\section{The open single-mode Jaynes-Cummings model}
	
	We first review the standard case of a single-mode cavity \cite{Jaynes1963a,Haroche2006} containing a two-level system, and derive Eqs.~(1)-(3) in the main text.
	The open single-mode Jaynes-Cummings model is given by the Master equation and Hamiltonian~\cite{Haroche2006}
	\begin{align}
		\dot{\rho} &= -i[H; \rho] + \frac{\kappa}{2} \left(2 \hat{a} \rho \hat{a}^\dagger - \hat{a}^\dagger \hat{a} \rho - \rho \hat{a}^\dagger \hat{a}\right) \,,\\
		H &= \omega_{1} \hat{a}^\dagger \hat{a} + \frac{\omega_\mathrm{a}}{2} \hat{\sigma}^z + \left[g^*  \hat{a} \hat{\sigma}^+ + h.c.\right] \,.
	\end{align}
	Here, $\omega_1$ is the cavity mode's resonance frequency, $\kappa$ is its decay rate with the bath assumed in the vacuum state, $\omega_\mathrm{a}$ is the transition frequency of the two-level atom, $g$ is the mode-atom coupling strength, and $\hat{a}$ ($\hat{\sigma}^{z/\pm}$) denotes the operator(s) associated with the cavity mode (two-level atom).	We note that additional direct loss channels from the atom could easily be added to the model, but we omit them here for simplicity. We use units of $\hbar=1$ for the Hamiltonian.
	
	The complex level shift of the atom is a weak coupling observable, where the cavity acts as a Markovian environment for the atom. In the above model, we can extract it by adiabatically eliminating the cavity mode (see, e.g.,~\cite{Heeg2013b}). The resulting atomic Master equation is given by
	\begin{align}
		\dot{\rho}_\mathrm{a} =& -i[H_\mathrm{adiab}; \rho_\mathrm{a}] \nonumber + \frac{\Gamma}{2} \left(2 \hat{\sigma}^- \rho_\mathrm{a} \hat{\sigma}^+ - \hat{\sigma}^+ \hat{\sigma}^- \rho_\mathrm{a} - \rho_\mathrm{a} \hat{\sigma}^+\hat{\sigma}^-\right) \,,\\
        H_\mathrm{adiab} =& \frac12(\omega_\mathrm{a}+\Delta) \hat{\sigma}^z \,.
	\end{align}
	The cavity-induced frequency shift $\Delta$ --- often referred to as Lamb shift in the cavity QED literature \cite{Limonov2017} --- and the Purcell enhanced line width $\Gamma$~\cite{Purcell1946} are then given by
	\begin{align}
		\delta := \Delta - i\frac{\Gamma}{2} = \frac{gg^*}{\omega_\textrm{a} - \omega_1 + i\kappa/2}\,.
	\end{align}
	Together, they form the complex level shift in Eq.~(\ref{main::eq::single_mode_parameters}) of the main text.
	
	In order to derive spectroscopic observables, one typically considers an external bath Hamiltonian of Gardiner-Collett form \cite{Gardiner1985}. The Heisenberg-Langevin equations of motion for the above model then read \cite{Gardiner2004}
	\begin{align}
	\dot{\hat{a}}(t) &= -i\left(\omega_1 - i\frac{\kappa}{2}\right)\hat{a}(t) - ig\hat{\sigma}^-(t) - 2\pi i \kappa_R \hat{b}_\mathrm{in}(t)\,,
	\\
	\dot{\hat{\sigma}}^-(t) &= -i\omega_\textrm{a} \hat{\sigma}^-(t) + ig^*\hat{\sigma}^z(t)\hat{a}(t) \,,
	\end{align}
	where $\kappa_R$ is the mode-bath coupling strength, and $\hat{b}_\mathrm{in}(t)$ the input operator of the bath. 
	The emitted radiation outside the cavity can then be calculated via the input-output relation~\cite{Gardiner1985}
	\begin{align}
	\hat{b}_\mathrm{out}(t) = \hat{b}_\mathrm{in}(t) - i\kappa_R \hat{a}(t) \,.
	\end{align}
	In the linear spectroscopy regime, we can \dl{generally approximate $\langle\hat{\sigma}^z(t)\hat{a}(t)\rangle \approx -\langle \hat{a}(t)\rangle$}, such that the equations of the first order expectation values form a closed system of linear coupled differential equations. The solution for the reflection spectrum defined by $ \langle \hat{b}_\mathrm{out}(\omega) \rangle = r(\omega) \langle \hat{b}_\mathrm{in}(\omega) \rangle$ is then obtained as \cite{Nunnenkamp2011,Lentrodt2020}
	\begin{align}
	r(\omega) = r_\mathrm{cav}(\omega) - 2\pi i \frac{\frac{|\kappa_Rg|^2}{(\omega - \omega_1 + i\kappa/2)^2}}{\omega - \omega_\textrm{a} - \frac{gg^*}{\omega - \omega_1 + i\kappa/2}} \,,
	\end{align}
	where the empty cavity reflection coefficient is
	\begin{align}
	r_\mathrm{cav}(\omega) = 1 - 2\pi i \frac{|\kappa_R|^2}{\omega - \omega_1 + i\kappa/2} \,.
	\end{align}
	Note that similar expressions for the scattering matrix are obtained if multiple external channels are considered~\cite{Lentrodt2020}, in case of which $\kappa_R$ acquires a channel index. \dl{In general, $\kappa$ and $\kappa_R$ are not directly related, since the former includes a sum over all external channels.} \dl{We note that this expression fully applies at strong coupling and defines the witness observable $\delta(\omega)$, whose imaginary part $\Gamma(\omega)$ is directly proportional to the bath spectral density.}
	
	At weak coupling ($\kappa \gg g$) and close to the atomic resonance ($\kappa \gg \omega-\omega_\textrm{a}$), we can approximate the cavity properties as constant on the scale of the light-matter interaction, such that $1/(\omega - \omega_1 + i\kappa/2)\approx1/(\omega_\textrm{a} - \omega_1 + i\kappa/2)$. The spectrum then separates into a constant cavity background and an atomic line as
	\begin{align}\label{eq::singleMode_refSpec_appendix}
	r(\omega) \approx r_\mathrm{cav}(\omega_\textrm{a}) - 2\pi i  \frac{(\kappa^{(\mathrm{int})}_R)^2}{\omega - \omega_\textrm{a} - \Delta + i\Gamma/2} \,,
	\end{align}
	which is Eq.~(2) in the main text. The resonance modulation depth is given by
	\begin{align}
	(\kappa^{(\mathrm{int})}_R)^2 = \frac{|\kappa_Rg|^2}{(\omega_\textrm{a} - \omega_1 + i\kappa/2)^2} \,.
	\end{align}
	The cavity induced Lamb shift $\Delta$ and Purcell enhanced line width $\Gamma$ can thus be extracted as line shape parameters from the linear reflection spectrum at weak coupling.
	
	\section{\label{sec:SIqom}Details on the pFM expansion and diagonalization}
	\subsection{Background}
	In the main text, we employed a recently-developed pFM model applicable to resonator QED~\cite{Medina2021} and connected it to other techniques to obtain a certification criterion for multi-mode effects. In the following, we provide details on the motivation for this particular approach as outlined in the introduction and how the different branches of the literature connect to each other.
	
	The central feature of the model in \cite{Medina2021} is that it describes electromagnetic resonators by a pFM Master equation which is exactly Markovian, that is it does not employ a Markov approximation. This aspect can be understood in two different ways. On the one hand, it constitutes a version of the pseudomodes approach \cite{Garraway1997a,Garraway1997b}, which has recently been generalized and formalized \cite{Tamascelli2018,Mascherpa2020,Pleasance2020,Trivedi2021,Pleasance2021_arxiv}. On the other hand, the resulting Master equation can be seen as the special case of a Gardiner-Collett Hamiltonian where the system-bath coupling is frequency-independent~\cite{Viviescas2003,Lentrodt2020,Medina2021}. The model by Medina et al.~\cite{Medina2021} assumes the existence of such a basis for the case of resonator QED and employs a fitting procedure to obtain the resulting Master equation parameters.

	The central feature of this approach is therefore the absence of a Markov approximation, which implies applicability in the regime of overlapping modes and large losses \cite{Barnett1988a,Dalton1999a,Dalton2001,Dutra2000,Hackenbroich2002,Lentrodt2020,Fregoni2022} --- a central motivation of our paper and a crucial requirement for various experimental platforms mentioned in the introduction, including x-ray cavity QED. On the other hand, a limitation is that current methods rely on fitting procedures \cite{Medina2021,Mascherpa2020}. The latter in turn imply practical restrictions and existence assumptions \cite{Tamascelli2018,Trivedi2021}.
	
	Interestingly, there is a completely alternative approach to the problem of constructing Master equations for resonator QED with a discrete set of modes, given by quantized quasinormal modes \cite{Franke2019}.  \dl{The latter features the opposite merits compared to \cite{Medina2021}, in that it does not require a fitting procedure, but relies on a Markov approximation \cite{Franke2019_supplement} and restriction to the QNM sector with positive real part of the complex pole frequency, where the latter ensures Hermiticity of the QNM operators' commutator matrix \cite{Franke2019}.} Quasinormal modes have also been studied extensively in the semi-classical and perturbative light-matter interaction literature \cite{Lalanne2018,Kristensen2020} and have proved to be a practically useful basis for understanding resonator structures. Consequently, a large tool box with regards to numerical solvers is available for this basis \cite{Lalanne2018}.
	
	Our criterion developed in the main text connects these two sides in that it certifies multi-mode effects of the exactly Markovian pFM models \cite{Medina2021} by their non-hermitian diagonalization, which is related to the \dl{classical} quasinormal modes expansion. Our approach is therefore not based on the quantized quasinormal modes \cite{Franke2019} as a basis, which do require a Markov approximation \cite{Franke2019_supplement}. Instead, it works within the exactly Markovian pFM basis and utilizes classical QNM expansions as a tool to circumvent fitting procedures.
	
	In the following, we detail the model setup and diagonalization procedure. The precise connection to quasinormal modes expansions is then outlined in Sec.~\ref{sec::quasi}.
	
	\subsection{Few-mode model setup}
	Following the discussion in~\cite{Medina2021}, the pFM model Hamiltonian is written as 
	\begin{align}
	H_\mathrm{cav} = \sum_{ij} \omega_{ij} \hat{a}_i^\dagger \hat{a}_j \,,\label{eq:genmodel}
	\end{align}
	where the mode interaction parameters $\omega_{ij}$ form a real symmetric matrix \cite{Medina2021} and can also be motivated as a basis-transformed version of cross-mode decay terms \cite{Hackenbroich2002,Viviescas2003,Lentrodt2020}.
	The pseudomodes concept imposing spectrally flat and independent baths then results in a Markovian Master equation \cite{Medina2021}
	\begin{align}
	\dot{\rho} = -i[H; \rho] + \sum_i \frac{\kappa_i}{2} \left(2 \hat{ a}_i \rho \hat{a}_i^\dagger - \hat{ a}_i^\dagger \hat{ a}_i \rho - \rho \hat{a}_i^\dagger\hat{ a}_i\right) \,,
	\end{align}
	with the real bath coupling parameters $\kappa_i$.
	The two-level system inside the cavity is governed by the light-matter interaction Hamiltonian \cite{Medina2021}
	\begin{align}
	H = H_\mathrm{cav} + \frac{\omega_\textrm{a}}{2}\hat{\sigma}^z + \sum_i \left[g^*_i  \hat{a}_i \hat{\sigma}^+ + h.c.\right] \,,
	\end{align}
	where we used the rotating wave approximation and set $\hbar=1$. We note that in \cite{Medina2021}, the $g_i$ are assumed to be real-valued. Here, we allow for the option of complex-valued coupling constants, to avoid implicit assumptions and since it provides additional insight into the appearance of multi-mode effects as discussed in Sec.~\ref{sec:SIMM}. We note that as in \cite{Medina2021}, our approach here assumes the existence of such a representation and subsequently constrains its properties via the introduced certification criterion.
	
	\subsection{Few-mode expansion}
	In the weak coupling limit, upon adiabatic elimination of the cavity modes similarly to the single-mode case above, the cavity-induced complex level shift can be written in terms of the few-mode expansion as
	\begin{align}
	\generallevshift(\omega_\mathrm{test})&= \underline{g}^\dagger\, \doubleunderline{G} \,\underline{g}\,, \label{eq:levshift-mode}\\[2ex]
	\doubleunderline{G}  &= \left[\omega_\mathrm{test}\mathbb{I}-\doubleunderline{\tilde{H}}_\mathrm{cav}\,\right]^{-1} \,,\\[2ex]
	\doubleunderline{\tilde{H}}_\mathrm{cav} &= \doubleunderline{\omega}_\mathrm{cav} - \frac{i}{2}\doubleunderline{\kappa} \,,
	\end{align}
	where the elements of the vectors and matrices are given by $(\underline{g})_i = g_i$, $(\doubleunderline{\omega}_\mathrm{cav})_{ij} = \omega_{ij}$ and $(\doubleunderline{\kappa})_{ij} = \kappa_i \delta_{ij}$.

	We note that $\doubleunderline{G}$ also appears in the expression for the spectral density in \cite{Medina2021}, where it was used to fit the model parameters. The spectral density is related to the imaginary part of the level shift. The Lamb shift, whose multi-mode features we investigate in the main text, is the real part of the level shift. We further note that we mainly use the latter as a witness observable for certifying properties of the pFM model, which is also applicable at strong coupling. At ultra-strong coupling, however, the rotating wave approximations both in the matter-cavity coupling and in the cavity losses have to be revisited \cite{DeLiberato2009,Bamba2014,Salmon2022}. \dl{Also at multi-mode strong coupling \cite{Krimer2014} or for wave-guide QED setups \cite{Johnson2019,Blaha2022}, a high degree of convergence of the witness observables over a large spectral range may be required \cite{Lentrodt2020}, since the Rabi splitting then bridges the free spectral range.}
	
	\subsection{Diagonalization}
	The interaction matrix $\doubleunderline{\tilde{H}}_\mathrm{cav}$ in the few-mode expansion can be cast into diagonal form via an invertible transformation matrix $\doubleunderline{V}$. Introducing
	\begin{align}
	(\doubleunderline{V}\,\underline{g})_i &= \tilde{g}_i\,,\\
	(\doubleunderline{V}^{-1}\,\underline{g}^*)_i &= \bar{g}^*_i\,,\\
	(\doubleunderline{V}\,\doubleunderline{\tilde{H}}_\mathrm{cav}\,\doubleunderline{V}^{-1})_{ij} &= (\tilde{\Omega}_i - i \frac{\tilde{\kappa}_i}{2})\delta_{ij}\,,
	\end{align}
	the few-mode expansion Eq.~\eqref{eq:levshift-mode} can be written as
	\begin{align}
	\generallevshift(\omega_\mathrm{test})= \underline{g}^\dagger\, \doubleunderline{G} \,\underline{g} = \sum_i \frac{\bar{g}_i^*\tilde{g}_i}{\omega_\mathrm{test} - \tilde{\Omega}_i + i \frac{\tilde{\kappa}_i}{2}} \,.
	\end{align}
	This expression is now of the same form as a Mittag-Leffler pole expansion (see also Eq.~\eqref{eq::ML_poleExpansion_final} below) and we can read off the poles and residues in terms of the diagonalized mode frequencies and couplings as
	\begin{align}
	\tilde{\omega}_{\mathrm{pole},i} &= \tilde{\Omega}_i - i \frac{\tilde{\kappa}_i}{2} \,, \label{eq:rel1}
	\\
	r_i &= \bar{g}_i^* \tilde{g}_i \,.\label{eq:rel2}
	\end{align}
	In the context of electromagnetic resonators, such pole expansions are also known as quasinormal mode expansions, which have been studied extensively in the semi-classical and perturbative literature \cite{Lalanne2018,Kristensen2020} and more recently on a quantized level \cite{Franke2019}, as discussed above. In the following supplementary section \ref{sec::quasi}, we provide details on the connection and why it is useful for determining the properties of the few-mode expansion without a fitting procedure.
	
	We note that the transformation $\doubleunderline{V}$ is typically not unitary for the lossy systems that we are considering, such that applying it on an operator level yields modified commutation relations and $\bar{g}_i$ is generally not equal to $\tilde{g}_i$. The pole expansion thus corresponds to a non-hermitian diagonal basis, whose associated field operators do not necessarily feature standard bosonic commutation relations (see \cite{Franke2019} for the related quantization of QNMs).
	
	The assumption of no exceptional points enters since otherwise, the few-mode interaction matrix $\doubleunderline{\tilde{H}}_\mathrm{cav}$ is not diagonalizable by an invertible matrix. On the pole expansion side, we incorporated the same assumption by only including simple pole terms, to achieve self-consistency.
	
	\section{Quasimode expansion of the complex level shift}\label{sec::quasi}
	\begin{figure}[t]
		\includegraphics[width=0.5\columnwidth]{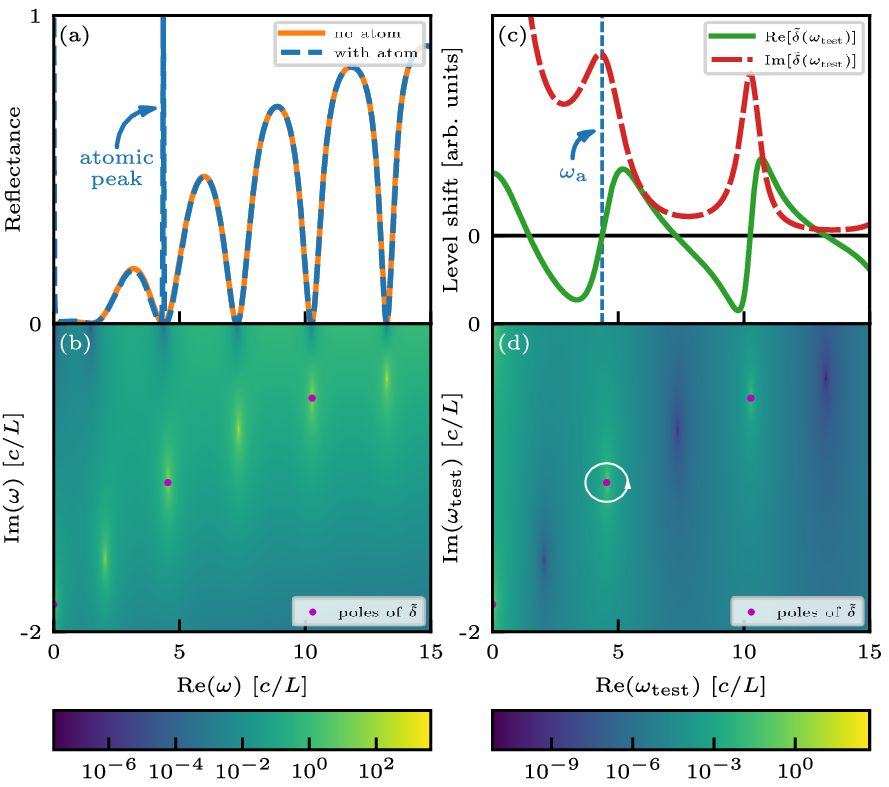}
		\caption{Illustration of the cQNM/pole expansion via the reflection spectrum and level shift in the complex frequency plane. (a) Reflectance with and without the atom. (b) Corresponding spectrum without the atom analytically continued to complex frequency plane. (c) Lamb shift and Purcell enhanced line width as a function of frequency. (d) Absolute value squared of the level shift in the complex frequency plane. Poles are marked in magenta. Note that the empty-cavity reflection has additional poles, which the atom does not couple to due to symmetry. An exemplary contour for the numerical determination of the pole residue is shown in (d) as a white circle. Further note that $n_\textrm{mirror}$ here and in the main text merely acts as a model parameter with unrealistically high values, which is used to simulate sufficiently good mirrors. In practice, highly reflecting mirrors instead comprise multi-layer dielectrics.\label{fig::illu_new_supp}}
	\end{figure}
	
	The cavity-induced energy shift and line broadening can alternatively be expressed via the classical electromagnetic Green's tensor $\mathbf{G}$ of the cavity environment as~\cite{Dung2000,Scheel2008,Asenjo-Garcia2017a}
	\begin{align}\label{eq::green_LS_linewidth}
	\levshift &= -\mu_0 \omega_\mathrm{a}^2\mathbf{d}^* \cdot \mathbf{G}(\mathbf{r}_\mathrm{a}, \mathbf{r}_\mathrm{a}, \omega_\mathrm{a}) \cdot \mathbf{d}\,, 
	\end{align}
	where $\mathbf{r}_\mathrm{a}$ is the atom's position and $\mathbf{d}$ its transition dipole moment. This expression applies rather generally for dipole transitions in absorptive dielectric environments \cite{Scheel2008}, but does not directly provide insight into the modal structure of the resonator, which is instead encoded in the pFM or quasinormal mode expansion (see main text), depending on which basis is required. Again, Eq.~\eqref{eq::green_LS_linewidth} is only valid at weak coupling, but for our purposes merely serves as a diagnostic observable to certify properties of the few-mode model, which also applies at strong coupling.
	
	The Green's function is useful in this context as it provides access to the pole expansion without the need for a fitting routine. To this end, we introduce a test frequency $\omega_\mathrm{test}$ and generalize the complex level shift to a frequency-dependent quantity (indicated by the tilde) as
	\begin{align}\label{eq::ML_levShiftTest}
		\generallevshift(\omega_\mathrm{test}) = -\frac{\mu_0 \omega_\mathrm{a}^2}{\hbar} \mathbf{d}^* \cdot \mathbf{G}(\mathbf{r}_a, \mathbf{r}_a, \omega_\mathrm{test}) \cdot \mathbf{d} \,.
	\end{align}
	Note that the $\omega^2_\mathrm{a}$ prefactor is kept constant, since we are only interested in characterizing the modal structure encoded in the Green's function. This choice is also important to achieve consistency with pFM models discussed in Sec.~\ref{sec:SIqom}, where the $g_i$-couplings implicitly depend on $\omega_\mathrm{a}$ in the same fashion \cite{Lentrodt2020}. Note, however, that this choice of coupling dependence in the pFM model may have to be adapted depending on the choice of gauge~\cite{FriskKockum2019}.
	
	The relevant cQNM expansion \cite{Lalanne2018,Kristensen2020} for our purposes is then given by \cite{Spiegel2009_book}
	\begin{align}
		 \generallevshift(\omega_\mathrm{test}) = \generallevshift(0) + \sum_i \left[ \frac{r_i}{\tilde{\omega}_{\mathrm{pole},i}} + \frac{r_i}{\omega_\mathrm{test} - \tilde{\omega}_{\mathrm{pole},i}} \right] \,,
	\end{align}
	where we again assume that the system does not contain exceptional points \cite{Miri2019,Ozdemir2019}, such that only simple poles are present.
	
	In practice, various numerical schemes for obtaining QNM expansions are available \cite{Lalanne2018}. A simple method is to find the poles $\tilde{\omega}_{\mathrm{pole},i}$ of the Green's function in the complex frequency plane and subsequently evaluate their residue $r_i$ by a line integral around the pole, as illustrated in Fig.~\ref{fig::illu_new_supp}(d)~\cite{Spiegel2009_book}. For the cases considered in the paper, we find that the constant term converges to zero, such that we can consider the simpler expansion
	\begin{align}\label{eq::ML_poleExpansion_final}
		\generallevshift(\omega_\mathrm{test}) = \sum_i \frac{r_i}{\omega_\mathrm{test} - \tilde{\omega}_{\mathrm{pole},i}} \,.
	\end{align}
	This property can also be understood as a general feature of the electromagnetic Green's function \cite{Defrance2020,Lalanne2018}.
	
	For more advanced numerical schemes, we refer to \cite{Lalanne2018,Wu2021a}. We note that one central achievement of our approach is that it connects pFM Master equations to cQNM expansions. The full breadth of techniques associated with the latter \cite{Lalanne2018,Kristensen2020} is therefore accessible and implies that our approach can be applied directly to a broad class of resonators that have previously been studied using quasinormal modes. We also note that both the pFM model and the Green's function expression apply in arbitrary dimensions.
	
	The physical level shift is then given by
	\begin{align}\label{eq::ML_poleExpansion_final_physical}
		\levshift = \generallevshift(\omega_\mathrm{a}) = \sum_i \frac{r_i}{\omega_\mathrm{a} - \tilde{\omega}_{\mathrm{pole},i}} \,.
	\end{align}
	We note that in the frequency-dependent generalized level shift $\generallevshift(\omega_\mathrm{a})$, the residues also acquire a dependence on $\omega_\mathrm{a}$, i.e., $r_i = r_i(\omega_\mathrm{a})$.
	
	A key feature of our approach is that the cQNM expansion is unique and independent of a basis choice. As we show, its truncation to a variable number of poles around $\omega_\mathrm{a}$ can then be used to certify and categorize multi-mode effects in the pFM model via the breakdown of characteristic features in the single-mode case discussed above. That is, by comparison of Eqs.~(\ref{main::eq::ML_poleExpansion_final_mainText}) and (\ref{main::eq::ML_poleExpansion_final_physical}), one can extract the complex pole frequencies $\tilde{\omega}_{\mathrm{pole},i} = \tilde{\Omega}_i - i \tilde{\kappa}_i/2$ and their residues $r_i = \bar{g}_i^* \tilde{g}_i$ of the diagonalized pFM expansion directly from the cQNM expansion, without the need for a fit. This identification can then in turn be used to draw conclusions about the pFM model, such as how many modes are required (see e.g.~Fig.~\ref{tab::classification}). This connection between cQNM and pFM forms the basis of our main results. In addition, these insights outline a path towards the construction of a backwards transformation (from the cQNM to the pFM basis), which remains an open problem~\cite{Pleasance2020}.
	
	An illustration of the reflection spectrum and the complex level shift in the complex frequency plane is given in Fig.~\ref{fig::illu_new_supp}, together with a contour line around a pole illustrating the numerical method for finding the pole expansion. The close correspondence between the poles in the complex level shift and the poles in the complex reflection spectrum is visible.

	\section{\label{sec:SIMM}Multi-mode effects in pFM models}
\begin{figure}[t]
		\includegraphics[width=0.8\columnwidth,trim=0 0 10 0]{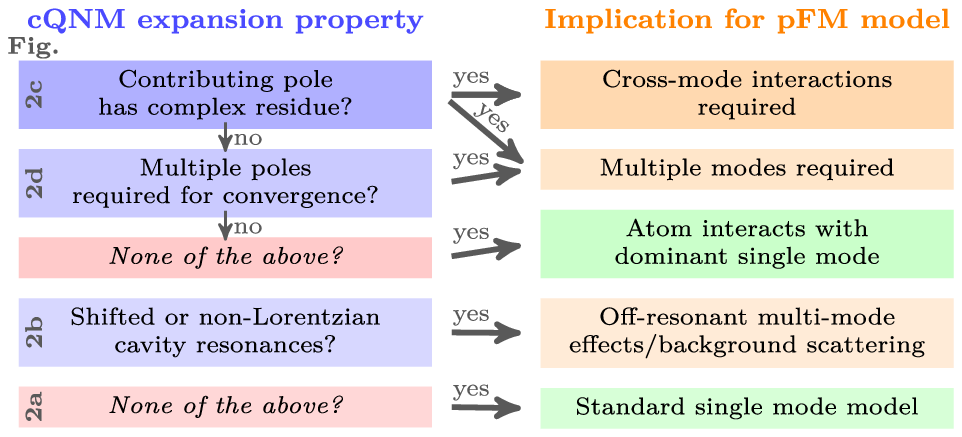}
		\caption{\label{tab::classification}Multi-mode certification criterion. The diagram shows which features of the classical quasinormal modes (cQNM) expansion [left column] can be checked to draw conclusions [arrows] about the pseudomodes-based few-mode (pFM) model [right column]. The resulting effect on the witness observable given by the Lamb shift is also depicted in Fig.~\ref{main::fig::illu_new2} in the main text. \dl{See also Fig.~\ref{main::fig::illu_new2}e in the main text.}}
	\end{figure}

	In this supplementary section, we provide details on the certification criterion for multi-mode effects. We explain how the QNM expansion provides insights into the structure of pFM models, in particular the number of contributing modes and whether they are interacting. The certification criterion is then given by observing the properties of the QNM expansion and using Fig.~\ref{tab::classification} to draw conclusions about the corresponding few-mode expansion. That is, we study the multi-mode properties of pFM Master equations via cQNM expansions.

	\subsection{Resonant effects}
	To this end, we specialize the derivation in Sec.~\ref{sec:SIqom} to the case of a diagonal interaction matrix $\omega_{ij}$ in Eq.~(\ref{eq:genmodel}). This case corresponds to the usual multi-mode extensions of the Jaynes-Cummings model, which do not consider direct interactions between the modes \cite{Medina2021,Franke2020a} or the equivalent cross-mode decay terms \cite{Hackenbroich2002,Viviescas2003,Lentrodt2020}. In this case, $\doubleunderline{\tilde{H}}_{\raisemath{2pt}{\mathrm{cav}}}$ naturally is of diagonal form, and the complex level shift reduces to
	\begin{align}
		\generallevshift^{(\mathrm{diag})}(\omega_\mathrm{test})= \sum_i \frac{\underline{g}^\dagger_i\underline{g}_i}{\omega_\mathrm{test} - \omega_{ii} + i \frac{\kappa_i}{2}} \,.
	\end{align}
	In this special case, the pFM expansion is therefore identical to a cQNM/pole expansion and the residues are \textit{real-valued}.

	\begin{figure}
		\centering
		\begin{minipage}[t]{0.45\textwidth}
			\centering
			\includegraphics[width=1.0\columnwidth]{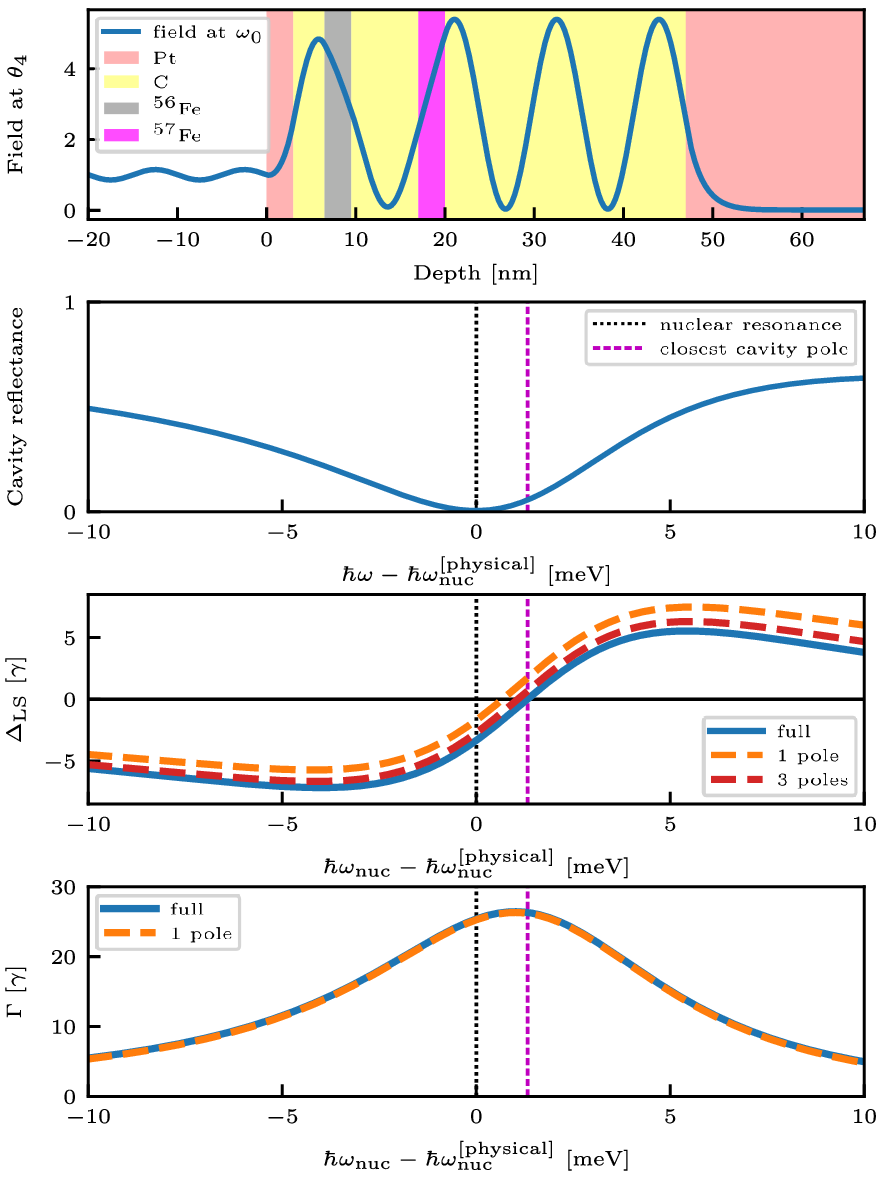}
			\caption{Multi-mode effects in the x-ray cavity considered in Fig.~4 of the main text. The x-ray incidence angle is chosen such that the fourth mode minimum lies on resonance. 
				(a) Cavity structure (see legend) and off-resonant cavity-field strength at the resonance frequency $\omega_\mathrm{nuc}^{[\mathrm{physical}]} = 14.4$~keV.
				(b) Off-resonant cavity reflectance as a function of energy. Note that the incidence angle is chosen such that the first reflection minimum lies on resonance. The real part of the main cavity pole is shifted with respect to the reflection minimum, indicating that this cavity features off-resonant multi-mode effects. 
				(c, d) cavity induced Lamb shift and Purcell enhanced line width, respectively, as a function of nuclear transition frequency relative to its physical value. We see that a pole expansion containing only the main pole already suffices to reproduce the line width, while few modes are required for convergence in the Lamb shift. Thus, the cavity does feature resonant multi-mode effects, in addition to the off-resonant ones. \label{fig::xray_cav_inv_mode4}}
		\end{minipage}\hfill
		\begin{minipage}[t]{0.45\textwidth}
			\centering
			\includegraphics[width=1.0\columnwidth]{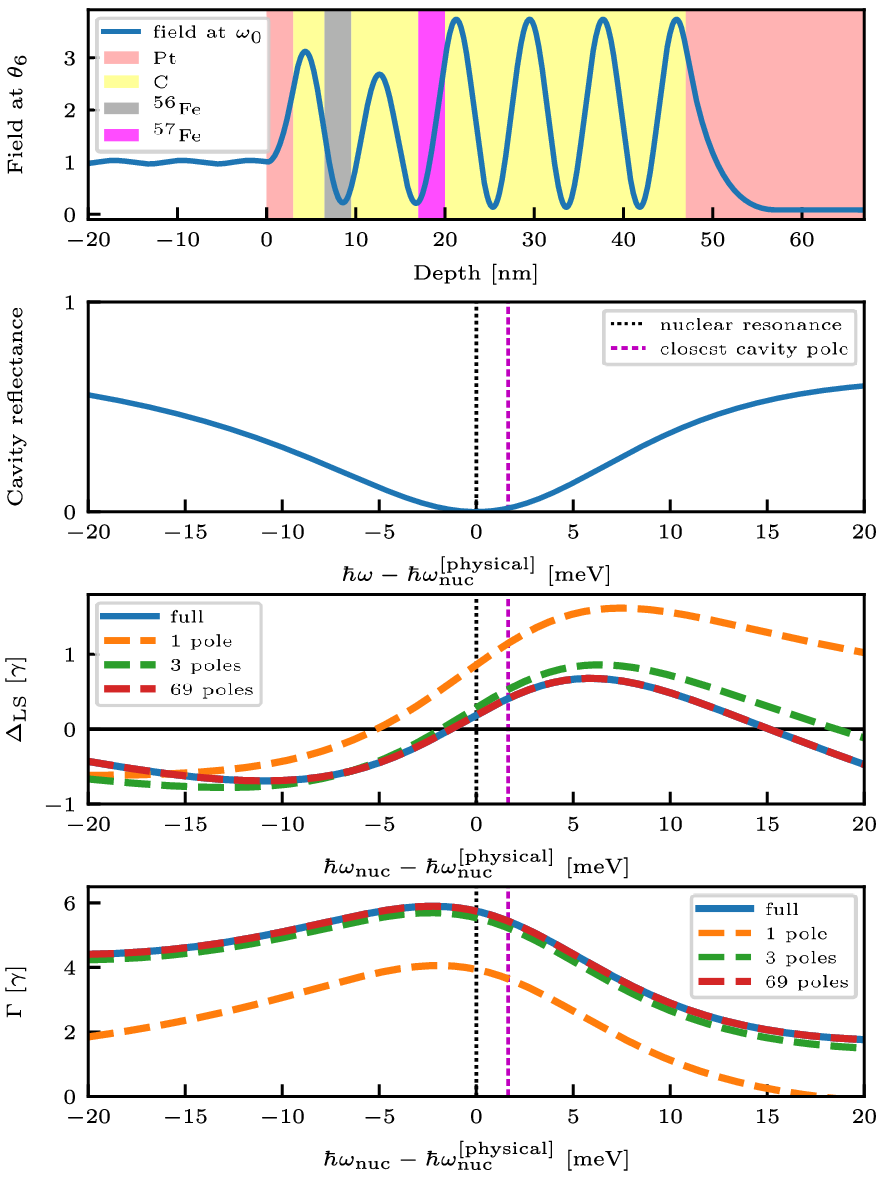}
			\caption{This figure is analogous to Fig.~\ref{fig::xray_cav_inv_mode4}, but with the x-ray incidence angle chosen such that the sixth mode minimum lies on resonance. In this case, we observe strong resonant multi-mode effects for both $\Delta$ and $\Gamma$, including multi-pole convergence and a significantly complex residue of the main pole. These results show that the inversion of the collective Lamb shift reported in Fig.~4 of the main text indeed is a multi-mode effect.\label{fig::xray_cav_inv_mode6}}
		\end{minipage}
	\end{figure}
	
	As a result, we can directly attribute the appearance of complex-valued residues in the quasinormal mode expansion to the non-diagonal elements of the mode coupling matrix $\omega_{ij}$ in the few-mode model, that is to \textit{interactions} between the modes. In terms of the few-mode theory, complex residues thus correspond to the case in which a single cavity degree of freedom (i.e., a single element of the non-hermitian diagonal basis) may be sufficient, albeit in the presence of significant interaction between multiple ``bare'' cavity modes. In the main text, we denoted this case as a \textit{complex residue effect}, which is illustrated in Fig.~\ref{main::fig::illu_new2} and Fig.~\ref{tab::classification}. In the single pole case, a complex residue effectively mixes the cavity-induced Lamb shift $\Delta$ and the superradiant enhancement of the line width $\Gamma$ in the complex level shift $\levshift$.
	
	A residue with a relevant imaginary contribution may therefore already appear in the case where a single pole is sufficient to achieve convergence in the QNM expansion. Further multi-mode effects may arise if multiple poles are required for the convergence of the QNM expansion, which we denote as \textit{multi-pole effects}. In this case, multiple cavity degrees of freedom (i.e., multiple elements of the non-hermitian diagonal basis) are required for convergence on the level of the few-mode model. Complex residues and the associated mode interactions may appear in combination with multiple poles.
	
	Both complex residue and multi-pole effects are \textit{multi-mode effects}, since they require multiple modes in the pFM expansion and cannot be explained by the standard single-mode Jaynes-Cummings model. \dl{We note that while the two effects have a distinct physical interpretation as outlined above, all physical examples provided in this work feature a combination of complex residue and multi-pole effects. It is currently unclear whether complex residues can appear practically without multiple poles contributing. However, one may suspect that such phenomena may appear in Fano cavities \cite{Denning2019,Cernotik2019} or related geometries, where the resonator properties can largely be designed.}
	
	This analysis shows that the features of the QNM expansion indeed certify multi-mode effects of light-matter interaction systems within pFM models. At the same time, they illustrate that the QNM and pFM basis are fundamentally different. The central difference is that the former is defined by its diagonalization property while the latter via the absence of a Markov approximation when deriving a Master equation in this basis~\cite{Medina2021}. This difference manifests in the categorization of multi-mode effects described above. For example, the complex residue effect implies that even for a system where a single pole/QNM is sufficient, one may require multiple modes in the pFM sense. The two bases therefore do not have a one-to-one mapping. The latter is also a central difference between the pFM Master equations to the quantized QNM Master equations constructed in \cite{Franke2019}. For the latter, one starts from a given number of QNM and finds a transformation to a Master equation with the same number of modes, using a Markov approximation \cite{Franke2019_supplement}. As our analysis of the inverse direction shows, such a transformation has a truncating character, which further supports that QNM are a different basis than the pFM basis.
	
	For completeness, we note that for many systems, poles at negative real frequency contribute to the QNM expansion\dl{, which are for example not included in the QNM quantization in \cite{Franke2019}, since the commutator matrix becomes non-Hermitian if these poles are included}. On the level of the \dl{classical} pole expansion, however, this feature does not make a difference. On the level of quantum optical models, a fully consistent treatment could potentially be to include counter-rotating terms in the system-bath coupling~\cite{Lentrodt2020}. These terms are omitted here, which is also connected to non-trivial existence assumptions of pFM models of the form suggested in \cite{Medina2021} (see also \cite{Tamascelli2018,Trivedi2021}). These issues, however, do not affect the classification and interpretation of the multi-mode effects in terms of the cQNM expansion as discussed in the main text, which crucially relies on identifying features that cannot possibly be explained without accounting for multiple modes and/or their interactions. \dl{In this sense, our method provides an exclusion criterion independent of the choice of basis which certifies that certain quantum models cannot be used to describe properties of the cQNM expansion, but also opens the door to investigate suitably generalized models.}
	
	\begin{figure}[t]
		\includegraphics[width=0.5\columnwidth]{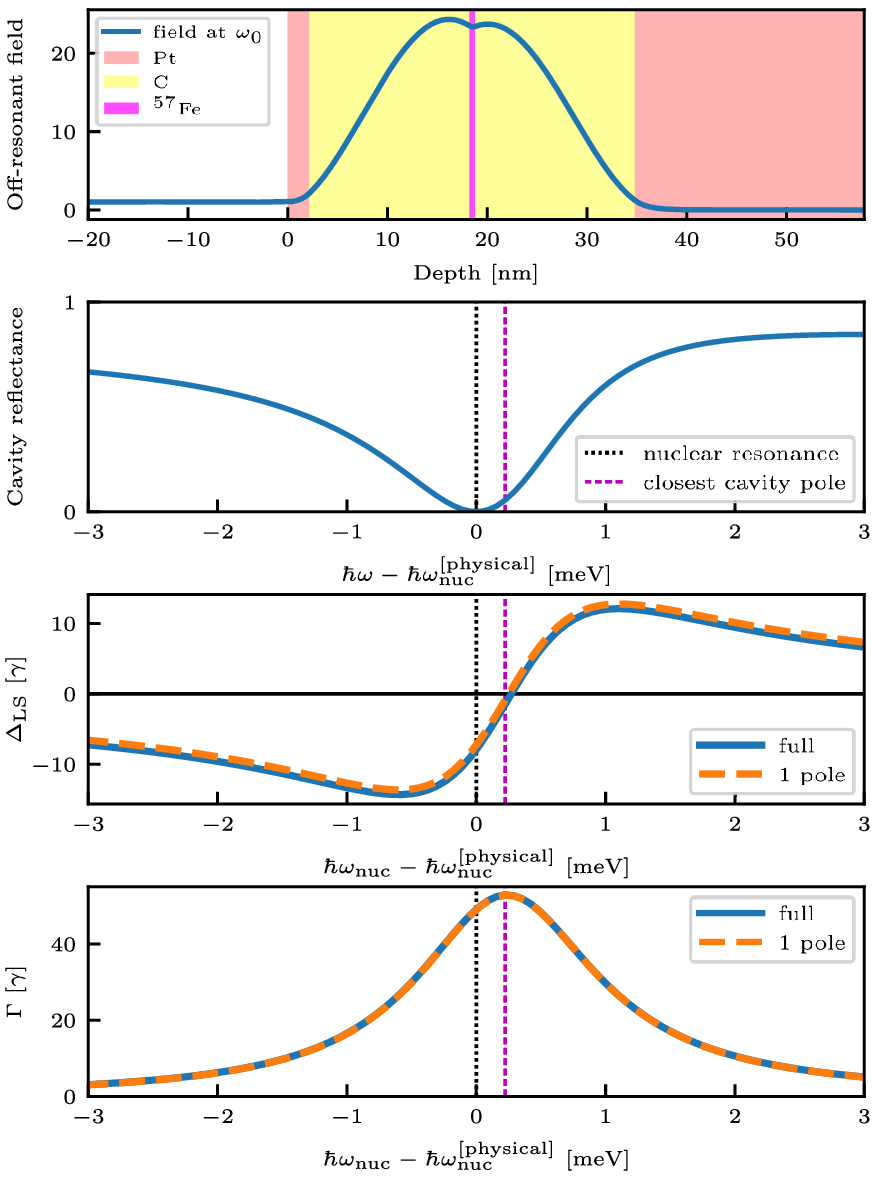}
		\caption{This figure is analogous to Fig.~\ref{fig::xray_cav_inv_mode4}, but for the cavity structure considered in the experiment reported in \cite{Rohlsberger2010}. We see that resonant multi-mode effects are negligible. The non-zero Lamb shift at the rocking minimum is explained by an off-resonant multi-mode shift.\label{fig::xray_cav_cls}}
	\end{figure}
	
	\subsection{Interpretation of off-resonant multi-mode effects in terms of background scattering}\label{sec::offres}
	For the above pFM model, one can in principle apply the input-output formalism \cite{Gardiner2004} to obtain spectroscopic observables such as the empty cavity reflection coefficient (that is without the atom). However, the result will still not be exact due to the truncation of the mode expansion. For isolated resonances, the latter can be a good representation of the scattering process. For the case of overlapping resonances, however, a background scattering contribution is required, as shown in \cite{Lentrodt2020}. In particular if a fitting procedure as the one in \cite{Medina2021} is employed, the bath modes will generally not correspond to the asymptotic scattering degrees of freedom.
	
	We note that this does not invalidate the approach for the purpose of computing the system dynamics in the cavity. Indeed, the input-output scattering can be translated to the full scattering result via the background scattering contribution \cite{Lentrodt2020} or the external field can be computed via direct field input-output relations once the atomic dynamics are known \cite{Medina2021}. Interestingly, since background scattering is an off-resonant cavity property, it can still be significant even if the few-mode expansion of the resonant dynamics is already well converged.
	
	In the present context, the\textit{ off-resonant multi-mode effects} can straightforwardly be identified as arising from non-negligible background scattering contributions. The interpretation as a separate phenomenon to the two resonant multi-mode effects therefore also directly maps to quantum optical few-mode models. Importantly, off-resonant effects imply that one may require a different number of modes to capture the scattering process than to solve the internal dynamics. This insight is particularly relevant for approaches based on fitting procedures of semi-classical scattering observables such as they are used in X-ray cavity QED \cite{Heeg2013b,Heeg2015c}. \dl{Also for the fitting procedures developed for pFM models \cite{Medina2021}, this observation shows that it is important to fit the resonant properties, such as the spectral density or Lamb shift, since scattering observables contain additional effects that are not captured in the Master equation, but instead contained in the background scattering contribution to the input-output relation.}

	\section{Multi-mode effects in x-ray cavity QED with M\"ossbauer nuclei}

	In the main text, we show that multi-mode effects can be used to invert the collective Lamb shift in x-ray cavity QED with M\"ossbauer nuclei. In the following, we illustrate this example further, and in particular show the contributions of the different multi-mode effects to the original Lamb shift experiment \cite{Rohlsberger2010} and the inversion effect. 
	
	Fig.~\ref{fig::xray_cav_inv_mode4} shows results for the cavity considered in the main text, with the x-ray angle of incidence chosen such that the reflection minimum of the fourth cavity mode is on resonance with the nuclei. Panel (a) shows the cavity structure together with the resulting field intensity, off-resonant with the nuclei. (b) shows the reflectance, which by design has a minimum at the  nuclear resonance. The panel further shows the position of the closest main cavity pole from the cQNM expansion. Its shift relative to the reflection minimum indicates the presence of off-resonant multi-mode effects. Panels (c,d) show the cavity-induced Lamb shift and the Purcell enhanced line width. It can be seen that the line width is already converged upon inclusion of the single main pole contribution. In contrast, a few poles are required for a converged Lamb shift. We thus find that the cavity also features resonant multi-mode effects, however, with only small line shape modifications, and less pronounced than the off-resonant multi-mode effects.
	
	Fig.~\ref{fig::xray_cav_inv_mode6} shows corresponding results for an x-ray incidence angle chosen such that the reflection minimum of the sixth mode is on resonance. In this case, in addition to an off-resonant multi-mode shift, resonant multi-mode effects occur. Both, the line width broadening and the cavity Lamb shift require the summation of multiple poles to achieve convergence. Furthermore, sizeable line-shape distortions appear, indicating the presence of complex-residue and multi-pole effects. The resonant multi-mode effects are strong enough to outweigh the off-resonant multi-mode effects, such that in total a reversed shift is obtained. 
	In particular, the imaginary part of the main pole's residue  (sixth pole, orange dashed line) contributes to the inversion. The contributions of higher modes act to reduce the positive shift again, but a significant shift remains at the minimum even when the result is fully converged.
	This clearly shows that the Lamb shift inversion can be attributed to resonant multi-mode effects.
	
	Finally, Fig.~\ref{fig::xray_cav_cls} summarizes the results of an analogous investigation for the cavity structure considered in the experiment reported in \cite{Rohlsberger2010}. It can be seen that resonant multi-mode effects are negligible in this case similarly to Fig.~\ref{fig::xray_cav_inv_mode4}. Instead, the non-zero Lamb shift at the rocking minimum, which is incompatible with single-mode models, is identified as an off-resonant multi-mode shift.
	
	\dl{We further note that despite the Lamb shift and Purcell enhanced line width being in principle related via Kramers-Kronig relations, our results (e.g.~Fig.~\ref{fig::xray_cav_inv_mode4}) show that the Lamb shift tends to converge more slowly than the line width. This suggests that for phenomenological approaches using fitting procedures, it may be advantageous to fit the Lamb shift instead of the spectral density as in \cite{Medina2021}, or even to consider both quantities to ensure full consistency.}

	\bibliographystyle{myprsty}
	\bibliography{library}